\documentclass[twocolumn,superscriptaddress,groupedaddress]{article}
\usepackage[a4paper, total={7.2in, 9in}]{geometry}
\usepackage{amsmath,bm,graphicx,epsfig,braket,amssymb,amsfonts,eufrak}
\usepackage{float}
\usepackage{multicol}
\usepackage{lipsum}
\usepackage{authblk}
\usepackage{color,colortbl}
\usepackage{booktabs}
\usepackage[table,dvipsnames]{xcolor}
\usepackage[labelfont=bf]{caption}
\usepackage{subcaption}
\usepackage[T1]{fontenc}
\usepackage{cite}
\usepackage{hyperref}
\hypersetup{
	 colorlinks=true,
	 linkcolor=blue,    
     urlcolor=blue,
	 citecolor=blue,
	 filecolor=cyan
}
\usepackage{array}
\usepackage{xr}

\begin{document}
	\title{\textbf{Computational Study Based Prediction of New Photocatalysts for water splitting by systematic manipulation of MXene surfaces.}}
	\author{Swati Shaw$^\ast$}
	\author{Subhradip Ghosh$^\dagger$}
	\affil{Department of Physics, Indian Institute of Technology Guwahati, Guwahati-781039, Assam, India.}
	\twocolumn[
	\begin{@twocolumnfalse}
		\maketitle
	\begin{abstract}
The compositional and structural flexibility of functionalised two-dimensional metal carbonitrides or MXenes has been exploited through a combinatorial search for new materials that can act as catalysts for photo-assisted water splitting by absorbing sunlight with energy in the infra-red region. Detailed calculations on 49 Janus MXenes where two surfaces are of asymmetric nature are carried out by first-principles Density Functional Theory. A screening procedure is adopted to arrive at potential candidates. Our calculations predict four new materials whose surfaces can activate both hydrogen and oxygen evolution reactions upon splitting water, two out of which are infra-red active, and the rest are visible light-active. We have performed a detailed microscopic analysis to find out the interrelations of the structural model of surface functionalisation, the chemistry of the surfaces, the electronic structure, and the alignment of bands with respect to the reaction potentials that explain our results. Apart from these four compounds, we find thirteen other compounds that are suitable for either hydrogen evolution or oxygen reduction reactions. This study lays out a guideline for the systematic discovery of potential new catalysts for water splitting under sunlight irradiation.
	\end{abstract}
\vspace{0.5cm}
\end{@twocolumnfalse}
]
	\section{Introduction}
	Photocatalysis is an advanced green technique that utilises solar energy to produce Hydrogen. Upon absorbing enough energy from sunlight, electrons and holes are generated and separated on the surfaces of a photocatalyst for further consumption in water splitting reactions: the Oxygen evolution reaction (OER) ( $ H_2O + 2 h^+ \longrightarrow \frac{1}{2}O_2 + 2H^+$) and the Hydrogen evolution reaction (HER) ( $2H^+ + 2e^- \longrightarrow H_2$). The minimum energy requirement for both reactions to happen simultaneously is 1.23 eV, with oxidation (reduction) potentials of O$_{2}$(H$_{2}$) are -5.67(-4.44) eV. This means that a photocatalyst for water splitting reactions must be a semiconductor possessing the following properties: (a) it must have a band gap of at least 1.23 eV but less than 3 eV so that it can capture a significant fraction of visible light which accounts for more than 40 \% of the solar energy compared to only 5  \%  in the ultra-violet spectrum, (b) the valence band maxima (VBM) and the conduction band minima (CBM) must lie below the oxidation potential of O$_{2}$ and above the reduction potential of H$_{2}$, respectively, (c) it should have small exciton binding energy and low recombination rate of charge carriers \cite{compsearch}.
	\begin{figure}[t]
		\centering
		\includegraphics[scale=0.4]{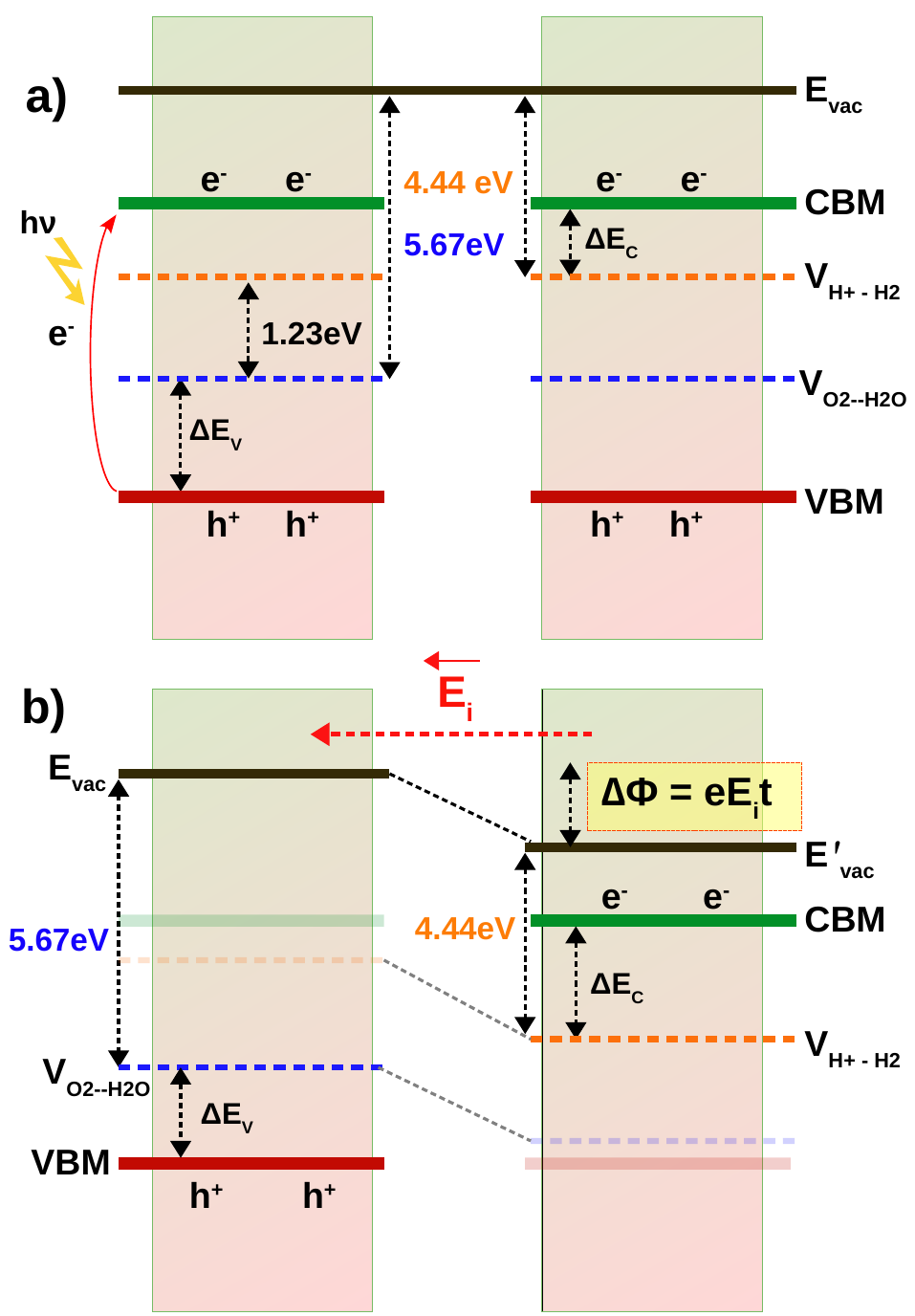}
		\caption {The Schematic Illustration of Photocatalytic Water Splitting from surfaces of a catalyst a) in the absence of an Electric Field, b) in the presence of an internal Electric Field E$_{i}$.  $\Delta \Phi$ is the potential difference generated between two surfaces of the material with thickness $t$. $\Delta E_{c}$ and $\Delta E_{v}$ are the overpotentials with respect to potentials for hydrogen evolution and oxygen evolution reactions, respectively. }
		\hrule
		\label{fig1}
	\end{figure}
	
	Two-dimensional (2D) materials are the sought-after ones as catalysts in photo-induced OER and HER. The high specific area in 2D materials promotes tremendous surface active sites to enhance photocatalytic reactions. Due to this, the recombination rate is less, too. Moreover, the passivation of their edges by getting functional groups attached to the surface dangling bonds minimizes reactions with ions in the solution. The structural flexibility in 2D materials also offers tunability in their electronic structures. Recent experiments with 2D materials and their heterostructures like Graphene and its derivatives, Graphite Carbon Nitrides, transition metal di-chalcogenides, black phosphorous and its derivatives demonstrate that they are quite useful in the photocatalytic splitting of water\cite{2dphoto,liu, yu, lin, wu, sang, li,zhuang}, HER in particular.
	
	The latest addition to this group is MXenes. MXenes, with chemical formula M$_{n+1}$X$_{n}$T$_{n+1} (n=1,2,3,...)$ where M is a transition metal element, X either Carbon or Nitrogen and T a surface passivating functional group, offers tremendous structural and compositional flexibility, ideal for tuning a number of functional properties and various applications\cite{mostafaei,li2020, he,zhang2023,yang2020,kan2020,bhardwaj,alwarappan,thenmozhi,MENG,yu5}. MXenes, derived by exfoliating their bulk counterparts, the MAX compounds, have functional groups T passivating the surfaces during the synthesis process. Though the standard synthesis route enabled only -O, -F, and -OH groups as T, recent experiments \cite{kamysbayev} demonstrated that MXene surfaces can be functionalised with other group 16 (S, Se) and group 17 (Cl, Br, I) elements too. As a result, the scope of investigating structure-property relations in MXenes has widened. Experimentally, few works on MXenes as photocatalysts for water splitting are available \cite{mxene,mxene1}. In those works, MXene has been used as a co-catalyst to accelerate either OER or HER on the surfaces of other 2D materials. First-principles computational modeling and simulations far outnumber the experiments and propose a number of M$_{2}$CT$_{2}$ MXenes as catalysts that can promote simultaneous HER and OER on its surfaces \cite{zhang,guo,m2co2,sc2ct2,tizr,hf2co1-xsx,cui}. However, in spite of various combinations of M and T, very few compounds had the correct alignments of VBM and CBM with respect to the oxidation and reduction potentials. All these MXenes investigated experimentally and through first-principles simulations are active in visible and ultra-violet parts of solar energy. This leaves out the scope to utilise about 50 \% of the solar energy for photocatalytic water splitting. This energy comes from the Infrared part of the spectrum. However, this requires lifting the restriction on the magnitude of the photocatalyst's band gap. A schematic map to achieve this, proposed by Li {\it et al} \cite{PRL}, is shown in Figure \ref{fig1}. In Figure \ref{fig1}a), schematics of photocatalytic water splitting from a system comprising two identical nanoscale surfaces under the usual model are shown. In this case, the energy levels corresponding to both oxidation and reduction potentials are located inside the band gap. When an electric field is applied between these two surfaces, the energy bands bend in the direction of the electric field, leading to a charge re-distribution with the system's valence band on one surface and conduction band on the other. If the electric field is an induced one due to an intrinsic dipole moment generated because of significant heterogeneity in the surfaces, the differences between the oxidation potential on one surface and the reduction potential on another surface reduce from 1.23 eV to $1.23- \Delta \Phi$ eV where $\Delta \Phi$ is the electrostatic potential difference associated with the induced electric field (Figure \ref{fig1} b)). The idea has been successfully implemented in h-BN functionalised by two elements having significant electronegative difference \cite{PRL}, and in M$_{2}$X$_{3}$ (M=Al, Ga, In; X=S, Se, Te) \cite{m2x3}. This model has been used to predict new photocatalysts in the Transition metal di-chalcogenide family by constructing Janus compounds MSX (M=Mo, W; X=Se, Te)from pristine MS$_{2}$ \cite{WSSe, yuan, MoSSe3}. These compounds exhibit good solar-to-hydrogen efficiencies. Derivation of Janus compounds from pristine M$_{2}$XT$_{2}$ MXenes can be done in various different ways as heterogeneity in the surfaces is possible through manipulation of M as well of T. In spite of the possibility, to our knowledge, there are two works on ordered Janus MXenes exploiting their potential as photocatalysts in water splitting reactions \cite{wong,zhang_Sc}. In this work, the Janus compounds proposed were based upon replacing one of the M surfaces of M$_{2}$CO$_{2}$ MXenes by another transition metal element M$^{\prime}$. From first-principles calculations, it was found that some of these compounds with band gaps smaller than 1.23 eV have their CBM and VBM adjusted against the water oxidation and reduction potentials due to the bending of bands and thus can be used as Infra-red active photocatalysts.
	
	Despite quite a few work on the prediction of new photocatalysts in the MXene family, both conventional and Janus, and with mixed success, there is no investigation that systematically explores the inter-relations of surface compositions, structural properties, electronic structure and alignments of bands with respect to characteristic potentials for water splitting reactions in MXene family of compounds such that predictors for potential photocatalyst can be formulated. Moreover, there is a void in terms of infra-red active photocatalysts in this family. In order to address both these issues, we have adopted a first-principles electronic structure calculations based combinatorial approach by construction of 47 Janus MXenes and putting in place a screening strategy for systematic exploration of possible new photocatalyst with emphasis on the ones active in the Infra-red region. The extensive study made predicts four new photocatalysts; among them, two are active in the Infrared part. Apart from the prediction of new materials, this study has demonstrated the potential stored in the Janus compounds of the MXene family as far as catalysts for OER and HER in the presence of sunlight are concerned. 
	\section{Computational Details}
	Density functional theory (DFT) based Plane-wave Pseudopotential method as implemented in Vienna ab initio simulation package (VASP)\cite{Vasp, Vasp1} is used for calculations of the electronic structures.The projected-augmented wave (PAW) \cite{PAW, PAW1} pseudopotentials are used throughout. To obtain accurate band gaps and electronic structures, the Heyd-Scuseria-Ernzerhof (HSE06)\cite{HSE1} functional is used to treat the exchange-correlation part of the Hamiltonian. Screening parameters of 0.2 \AA$^{-1}$ and mixing exchange parameters of 0.25 are used in our calculations. The kinetic energy cut-off is set to 550 eV. To avoid artificial interactions between periodic images, a vacuum slab of 20\AA \hspace{0.2cm}is introduced parallel to the monolayers. For structural relaxation, the convergence criteria of total energy and the Hellmann-Feynman forces are set as 10$^{-6}$ eV and 10$^{-3}$ eV/\AA, respectively. For structural optimization and densities of states calculations, the Brillouin zones are sampled by 9$\times$9$\times$1 and 35$\times$35$\times$1 $\Gamma$-centered $k$-meshes. Due to the asymmetric structures of our materials, a dipole correction along the normal direction of monolayers is adopted in every calculation.
	\section{Results and Discussions}
	In the quest for new materials suitable for photocatalytic water splitting, we have considered 32 M$_{2}$CTT$^{\prime}$ and 15 MM$^{\prime}$CO$_{2}$ Janus MXenes and adopted the following screening procedure to pin down the promising materials. The procedure has the following steps: (1) The electronic structures of all compounds are calculated. The ones showing semiconducting behavior are considered for further processing; (2) for each one of the materials screened, the alignments of their band-edges, the internal electric field, and the overpotentials are investigated; and (3) the chemical nature of the band decomposed charge density distributions of the compounds showing appropriate band-edge alignments in step (2) are further investigated. The compounds having VBM and CBM contributions coming from two different surfaces, along with appropriate band alignments and overpotentials, are considered promising catalysts for photocatalytic water splitting. In the following section, we first discuss the structural models of surface-passivated MXenes. Results pertaining to steps (1)-(3) mentioned above are then presented.
	\begin{figure}[t]
		\centering
		\includegraphics[scale=0.36]{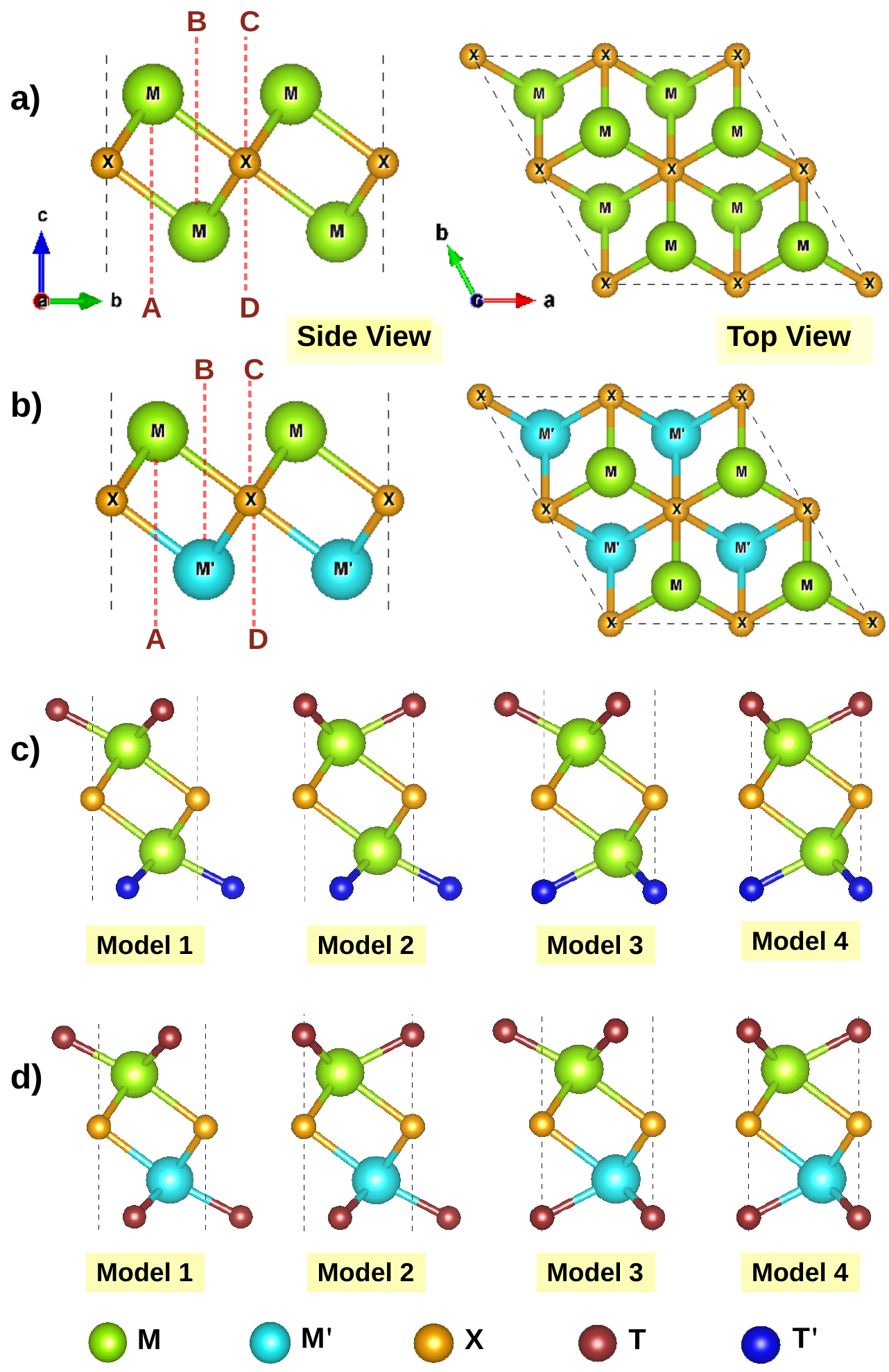}
		\caption {Side and top views with possible sites of functionalisation for a) M$_{2}$X and b) MM$^{\prime}$X MXene Surfaces. Side view of Four possible structural models of functionalisation in Janus MXenes c)  M$_{2}$XTT$^{\prime}$  d) MM$^{\prime}$XT$_{2}$ are shown.}
		\hrule
		\label{fig2}
	\end{figure}
	\subsection{Structural Models of functionalised MXenes}
	Pristine MXenes with chemical formula M$_{2}$C have hexagonal symmetry with space group P6\textsubscript{3}/mmc (Space group 194). Upon surface passivation with identical functional group T, symmetry of M$_{2}$CT$_{2}$ is reduced, making them members of P$\bar{3}$m1 space group (Space group 164). The making of Janus MXenes MM$^{\prime}$CT$_{2}$ and M$_{2}$CTT$^{\prime}$ breaks the inversion symmetry of  P$\bar{3}$m1. Consequently, the Janus MXenes belong to P3m1 symmetry (Space group 156). There is more than one choice for the surface functional group with regard to the site of passivation shown in Figure \ref{fig2}. There are altogether four sites, two hollow sites associated with the transition metal component (A and B in Figure \ref{fig2}a),b)) and two with the C atom (C and D in Figure \ref{fig2}a),b)). Accordingly, there can be four possible structural models of surface functionalised Janus MXenes: Model 1, where the functional groups occupy sites A and B; Model 2, where they occupy sites C and A; Model 3 where they occupy sites B and D and Model 4 where the sites of occupancy are C and D (Figure \ref{fig2}c),d)).
	For each system considered in this work, we have calculated the total energies of all four models. The structural model with the lowest energy is considered the ground state structure for a given compound.
	\subsection{Structural and Electronic properties}
	In this section, we systematically discuss the structural and electronic properties of three different sets of Janus MXenes. In each case, we analyse the results to understand the origin of these properties. In a combinatorial study, such insights are necessary for future studies on a larger scale.
	\subsubsection{ Janus M$_{2}$CTT$^{\prime}$ (M = Sc, Ti, Zr, Hf; T/T$^{\prime}$= H, F, OH, O) MXenes}
	Structural information like the structural model, the lattice constants, bond lengths, and thickness of compounds of this series are presented in Table \ref{table1}. We find that all 24 compounds investigated stabilise in Model 1 of functionalisation. This can be understood in the following way: To fulfill the octet rule, C and O require 4 and 2 electrons, respectively, whereas the requirement of  H, F, and (OH) is 1. All four transition metals have at least 3 valence electrons that can be used to form M-C, M-T, and M-T$^{\prime}$ bonds. Thus, the functional groups prefer to occupy hollow sites that are associated with the M sites stabilising them in Model 1.
	\begin{figure*}[t]
		\includegraphics[scale=0.29]{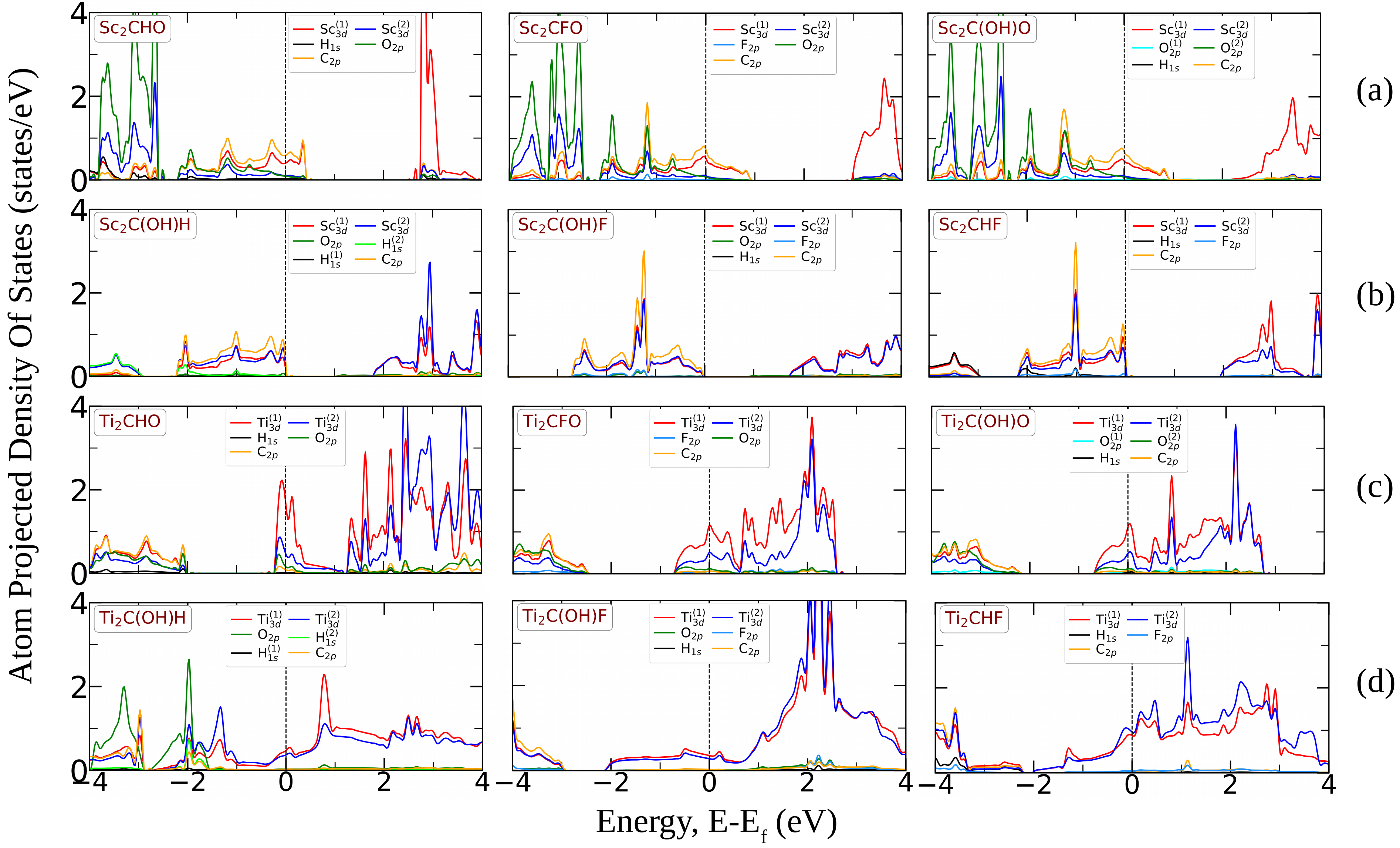}
		\caption{Atom projected densities of states of a) Sc\textsubscript{2}CTO, T = H, F, OH; b) Sc\textsubscript{2}CTT$^{\prime}$, T,T$^{\prime}$ = H, F, OH, c) Ti\textsubscript{2}CTO, T = H, F, OH ; d) Ti\textsubscript{2}CTT$^{\prime}$, T,T$^{\prime}$ = H, F, OH MXenes. The Fermi level in each case is set to 0 eV.}
		\label{fig3}
		\hrule
	\end{figure*}
	\begin{table*}[htb!]
		\small
		\caption{\bf{The structural parameters and electronic ground states of M$_{2}$CTT$^{\prime}$(M = Sc, Ti, Zr, Hf; T,T$^{\prime}$= O, OH, F, H) MXenes}}
		\label{table1}
		\begin{tabular*}{\textwidth}{@{\extracolsep{\fill}}ccccccccc}
			\hline
			\multicolumn{1}{c}{\textbf{Systems}} & \multicolumn{1}{c}{\textbf{Electronic}} & \multicolumn{1}{c}{\textbf{Stable}} & \multicolumn{1}{c}{\textbf{Lattice-Constant}} & \multicolumn{4}{c}{\textbf{Bond-Lengths }} & \multicolumn{1}{c}{\textbf{Thickness}} \\
			\cline{5-8}
			\bf{(M\textsubscript{2}CTT$^{\prime}$)} &\bf{ground state} & \bf{Configuration} & \bf{a=b}& \textbf{M-H} & \textbf{M-F} & \textbf{M-OH} & \textbf{M-O} & \bf{t}\\
			& & & (\AA) & (\AA) & (\AA) & (\AA) & (\AA) & (\AA)\\
			\hline
			Sc\textsubscript{2}CHO & Metal & Model 1 & 3.27 & 2.14 & - & - & 2.03 & 4.60\\
			Sc\textsubscript{2}CFO & Metal & Model 1 & 3.25 & - & 2.21 & - & 2.02 &  4.87 \\
			Sc\textsubscript{2}C(OH)O & Metal & Model 1 & 3.25 & - & - & 2.26 & 2.03 & 5.93 \\
			Sc\textsubscript{2}C(OH)H & Semiconductor & Model 1 & 3.30 & 2.17 & - & 2.27 & - & 5.76 \\
			Sc\textsubscript{2}C(OH)F & Semiconductor & Model 1 & 3.29 & - & 2.23 & 2.27 & - & 5.89 \\
			Sc\textsubscript{2}CHF & Semiconductor & Model 1 & 3.29 & 2.16 & 2.22 & - & - & 4.69 \\
			\\
			Ti\textsubscript{2}CHO & Metal & Model 1 & 3.02 & 2.01 & - & - & 1.98 & 4.42\\
			Ti\textsubscript{2}CFO & Metal & Model 1 & 3.03 & - & 2.15 & - & 1.98 & 4.67\\
			Ti\textsubscript{2}C(OH)O & Metal & Model 1 & 3.04 & - & - & 2.17 & 1.99 & 5.66 \\
			Ti\textsubscript{2}C(OH)H & Metal & Model 1 & 3.05 & 2.01 & - & 2.17 & - & 5.52 \\
			Ti\textsubscript{2}C(OH)F & Metal & Model 1 & 3.07 & - & 2.18 & 2.18 & - & 5.78 \\
			Ti\textsubscript{2}CHF & Metal & Model 1 & 3.04 & 2.00 & 2.16 & - & - & 4.52\\
			\\
			Zr\textsubscript{2}CHO & Metal & Model 1 & 3.29 & 2.17 & - & - & 2.13 & 4.67 \\
			Zr\textsubscript{2}CFO & Metal & Model 1 & 3.29 & - & 2.31 & - & 2.13 & 4.94 \\
			Zr\textsubscript{2}C(OH)O & Metal & Model 1 & 3.30 & - & - & 2.34 & 2.14 & 5.95\\
			Zr\textsubscript{2}C(OH)H & Metal & Model 1 & 3.31 & 2.16 & - & 2.34 & - & 5.84 \\
			Zr\textsubscript{2}C(OH)F & Metal & Model 1 & 3.32 & - & 2.34 & 2.35 & - & 6.15\\
			Zr\textsubscript{2}CHF & Metal & Model 1 & 3.30 & 2.16 & 2.32 & - & - & 4.83\\
			\\
			Hf\textsubscript{2}CHO & Metal & Model 1 & 3.25 & 2.14 & - & - & 2.11 & 4.64 \\
			Hf\textsubscript{2}CFO & Metal & Model 1 & 3.25 & - & 2.30 & - & 2.11 & 4.89\\
			Hf\textsubscript{2}C(OH)O & Metal & Model 1 & 3.26 & - & - & 2.30 & 2.12 & 5.88 \\
			Hf\textsubscript{2}C(OH)H & Metal & Model 1 & 3.26 & 2.13 & - & 2.31 & - & 5.77 \\
			Hf\textsubscript{2}C(OH)F & Metal & Model 1 & 3.28 & - & 2.32 & 2.32 & - & 6.07 \\
			Hf\textsubscript{2}CHF & Metal & Model 1 & 3.26 & 2.13 & 2.30 & - & - & 4.78 \\
			\hline
		\end{tabular*}
	\end{table*}
	
	The electronic ground states of the compounds in this series are compiled in Table \ref{table1}. We find that apart from three Sc-based Janus, all have metallic ground states. To understand the reasons, we take recourse to their bonding pictures and atom projected densities of states (Figure \ref{fig3} \& Figure S1, supplementary information). Janus Sc$_{2}$CHO, Sc$_{2}$CFO and Sc$_{2}$C(OH)O are obtained from Sc$_{2}$CO$_{2}$ by replacing the O passivating the top surface (the surface of Sc$^{(1)}$) with other functional groups that have 1 electron more than O. Sc$^{(2)}$, the Sc atom in the bottom surface, provides 2 electrons to the O passivating the surface and the remaining 1 to C. Sc$^{(1)}$, on the other hand, after fulfilling the requirement of H, F or (OH) passivating its surface by providing 1 electron, supplies remaining 2 to C. As a result, there is 1 unpaired in these systems associated with Sc$^{(1)}$-3$d$ C-$2p$ hybridisation. This is responsible for the metallic ground states in these compounds and is reflected in the features of their electronic structures near the Fermi level, where the states are dominated by Sc$^{(1)}$-$d$ and C-$p$ orbitals (Figure \ref{fig3} a)). When the Janus is formed by replacing the O passivating the surface of Sc$^{(2)}$ by H, F, or (OH), no unpaired electron remains in the system. This is responsible for the semiconducting ground states in Sc$_{2}$C(OH)H, Sc$_{2}$C(OH)F and Sc$_{2}$CHF. The atom projected densities of states of these compounds (Figure \ref{fig3} b)) corroborate this where the states near the Fermi level are now primarily made up of both Sc atoms and C.
	
	The origin of metallic ground states in Ti, Zr, and Hf-based Janus considered in this series can be understood in a similar way. All three transition metals have 1 electron more than Sc. In the Ti series, the compounds Ti$_{2}$CHO, Ti$_{2}$CFO and Ti$_{2}$(OH)O have the O passivating the Ti$^{(2)}$ surface. The 4 electrons of Ti$^{(2)}$ are provided to C and O while 1 less electron is required to be provided by Ti$^{(1)}$ as its surface is passivated by H, F, or (OH). This extra electron in the system is responsible for the metallic nature of these three compounds. The electronic structure of these compounds (Figure \ref{fig3}c)) reflects this as the states near the Fermi level are contributed mostly by Ti$^{(1)}$. When neither the surfaces are passivated by O, as is the case of Ti$_{2}$C(OH)H, Ti$_{2}$C(OH)F and Ti$_{2}$CHF, both Ti$^{(1)}$ and Ti$^{(2)}$ have one electron each extra. The states near the Fermi level of these compounds are thus dominated equally by the states from Ti on both surfaces (Figure \ref{fig3}d)). The electronic structures of compounds in the Zr and Hf series (Figure S1, supplementary information) can be explained in a similar manner. 
	\begin{figure*}[htb!]
		\includegraphics[scale=0.29]{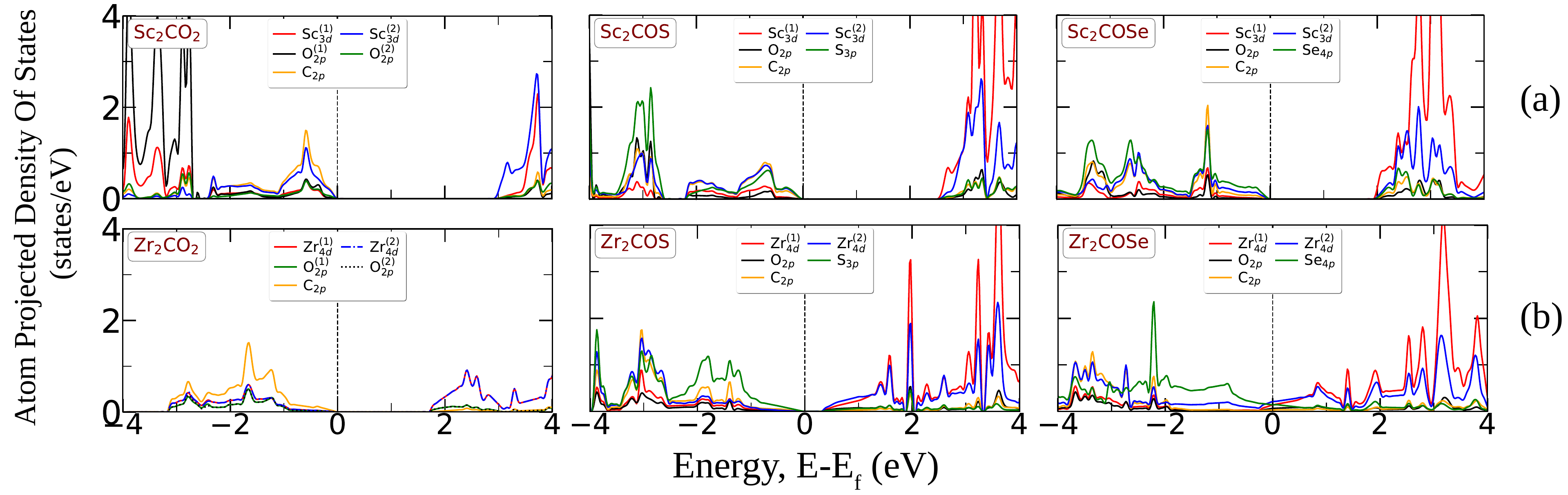}
		\caption{Atom projected densities of states for a) Sc\textsubscript{2}COS , Sc\textsubscript{2}COSe \& Sc\textsubscript{2}CO\textsubscript{2}, b) Zr\textsubscript{2}COS, Zr\textsubscript{2}COSe \& Zr\textsubscript{2}CO\textsubscript{2}. Fermi energy in each case is set to 0 eV.}
		\label{fig4}
	\end{figure*}
	\begin{table*}[htb!]
		\small
		\caption{\bf{The structural parameters and electronic ground states of M\textsubscript{2}COT$^{\prime}$ (M=Sc,Ti, Zr, Hf and T$^{\prime}$= S, Se) MXenes }}
		\label{table2}
		\begin{tabular*}{\textwidth}{@{\extracolsep{\fill}}ccccccc}
			\hline
			\multicolumn{1}{c}{\textbf{Systems}} & \multicolumn{1}{c}{\textbf{Electronic}} & \multicolumn{1}{c}{\textbf{Stable}} & \multicolumn{1}{c}{\textbf{Lattice-Constant}} & \multicolumn{2}{c}{\textbf{Bond-Lengths }} & \multicolumn{1}{c}{\textbf{Thickness}} \\
			\cline{5-6}
			\bf{(M\textsubscript{2}COT$^{\prime}$)} & \bf{ground state} & \bf{Configuration} & \bf{a=b}& \textbf{M-O} & \textbf{M-T$^{\prime}$} & \bf{t}\\
			& & & (\AA) & (\AA) & (\AA) & (\AA) \\
			\hline
			Sc\textsubscript{2}COS & Semiconductor & Model 3 & 3.63 & 2.15 & 2.46 & 4.03 \\
			Sc\textsubscript{2}COSe & Semiconductor & Model 3 & 3.66 & 2.16 & 2.58 & 4.16 \\
			\\
			Ti\textsubscript{2}COS & Metal & Model 1 & 3.11 & 1.99 & 2.40 & 4.99\\
			Ti\textsubscript{2}COSe & Metal & Model 1 & 3.12 & 2.00 & 2.56 & 5.17\\
			\\
			Zr\textsubscript{2}COS & Semiconductor & Model 1 & 3.38 & 2.14 & 2.53 & 5.22\\
			Zr\textsubscript{2}COSe & Metal & Model 1 & 3.38 & 2.15 & 2.69 & 5.45 \\
			\\
			Hf\textsubscript{2}COS & Semiconductor & Model 1 & 3.34 & 2.12 & 2.50 & 5.18 \\
			Hf\textsubscript{2}COSe & Metal & Model 1 & 3.34 & 2.13 & 2.66 & 5.38\\
			\hline
		\end{tabular*}
	\end{table*}
	\subsubsection{The series M\textsubscript{2}COT$^{\prime}$ (M=Sc,Ti,Zr,Hf; T$^{\prime}$ = S, Se)}
	In the previous sub-section, we have found that the absence of unpaired electrons in Janus MXenes is the reason behind the semiconducting ground state in certain members of the series. Since the Janus MXenes considered in the previous sub-section are created by replacement of  O atoms in M$_{2}$CO$_{2}$ MXenes, compounds that are all semiconductors \cite{guo, zhang,m2co2}, our next set of Janus compounds are the ones obtained by replacing one of the O by other group 16 elements S and Se in M$_{2}$CO$_{2}$. S and Se are isoelectronic to O; hence, it is expected that band filling pattern in M$_{2}$COS(M$_{2}$COSe) will be similar to M$_{2}$CO$_{2}$ and more Janus semiconductors will be produced.
	Unlike the compounds in the series M$_{2}$CTT$^{\prime}$ discussed in the previous sub-section, the structural model of surface passivation, compounds in this series stabilise in, vary (Table \ref{table2}). Sc-based compounds stabilise in Model 3, the one where T$^{\prime}$ functional group, instead of occupying the hollow site of the transition metal (site B, Figure \ref{fig2}) like O does, occupies the hollow site associated with C atom (site D, Figure \ref{fig2}). Each Sc, with only 3 valence electrons, is unable to provide 4 electrons required by C and the functional groups. Therefore, even in Sc$_{2}$CO$_{2}$, one of the O does not have a sufficient number of electrons to satisfy the octet rule and prefers to occupy the hollow site of C so that it can share the electrons of C. This is the reason why Sc$_{2}$CO$_{2}$ minimise total energy in Model 3 where Sc$^{(1)}$(Sc$^{(2)}$) surface is passivated by O occupying B(D) site. Consequently, Sc$_{2}$CO$_{2}$ is a semiconductor as seen from its atom projected densities of states (Figure \ref{fig4}a)). Due to different positions of passivation, the chemical environment of the two O atoms is different. As a result, O$^{(1)}$ and O$^{(2)}$ states occupy different parts of the occupied spectrum. Hybridisations are also influenced by this. Near the Fermi level, strong hybridisation is found between $p$ states of O$^{(2)}$, $p$ states of C, and $d$ states of both Sc atoms. Hybridisations of C-$p$ and O$^{(2)}$-$p$ orbitals push the unoccupied states at high energies, opening up a gap of $\sim$ 2 eV, making the ground state of Sc$_{2}$CO$_{2}$ semiconducting. In Sc$_{2}$COS(Sc$_{2}$COSe), the S(Se) replacing O$^{(2)}$ also prefers to occupy the hollow site of C for the same reason O$^{(2)}$ occupied it in Sc$_{2}$CO$_{2}$. The electronic structure of these two compounds is very similar to Sc$_{2}$CO$_{2}$. In both compounds, states near Fermi level in the occupied part see strong hybridisation between Sc$^{(2)}$-$d$ C-$p$ and S/Se-$p$ states. The band gap gradually decreases from Sc$_{2}$CO$_{2}$ to Sc$_{2}$COSe as the anti-bonding states gradually shift towards lower energy with an increase in the size of T$^{\prime}$ functional group. The $p$ states of T$^{\prime}$ are more delocalised as we go down the rows of group 16. Their hybridisations with other atoms, therefore, fill up states at lower energies in the conduction band and reduce the band gap.
	
	Ti, Zr, Hf based compounds in this series stabilise in Model 1 as C and functional groups receive the right number of electrons to satisfy the octet rule. But, their electronic ground states (Table \ref{table2}) differ considerably. The differences in their atom-projected densities of states (Figure \ref{fig4}b) and Figure S2, supplementary information) explain this. The electronic structures of three M$_{2}$CO$_{2}$ compounds with M being Ti, Zr, and Hf are very similar in terms of the identities of the orbitals hybridising strongly resulting in states near the valence and conduction band edges. The difference is in the positions of the bands and the edges, resulting in sizeable differences between their band gaps. While Zr$_{2}$CO$_{2}$ and Hf$_{2}$CO$_{2}$ have band gaps $\sim$ 2 eV, band gap in Ti$_{2}$CO$_{2}$ is only $\sim$ 1 eV. When one O is replaced with S or Se, there is a significant impact in the unoccupied part of the spectrum. The environment around the two surfaces becomes different. This affects the electronic structures of individual atoms. Notably, the densities of states of two M atoms are now different. States of M$^{(2)}$, the atom having S and Se on its surface, due to hybridisation with them, become substantially delocalised, forming the bottom of the conduction band and reducing the band gap considerably. Since the band gap of Ti$_{2}$CO$_{2}$ is smaller, such delocalisation gives rise to states in the gap when one O is replaced with S and Se. As a result Ti$_{2}$COS and Ti$_{2}$COSe are semi-metals. The reason Zr$_{2}$COS, Hf$_{2}$COS are semiconductors while Zr$_{2}$COSe, Hf$_{2}$COSe are metals is the degree of delocalisation of Se-hybridised states in comparison to S-hybridised states as observed from the atom-projected densities of states.  
	
	\subsubsection{The series MM$^{\prime}$CO\textsubscript{2}; M,M$^{\prime}$= Sc, Ti, Zr, Hf}
\begin{figure*}[htb!]
		\includegraphics[scale=0.28]{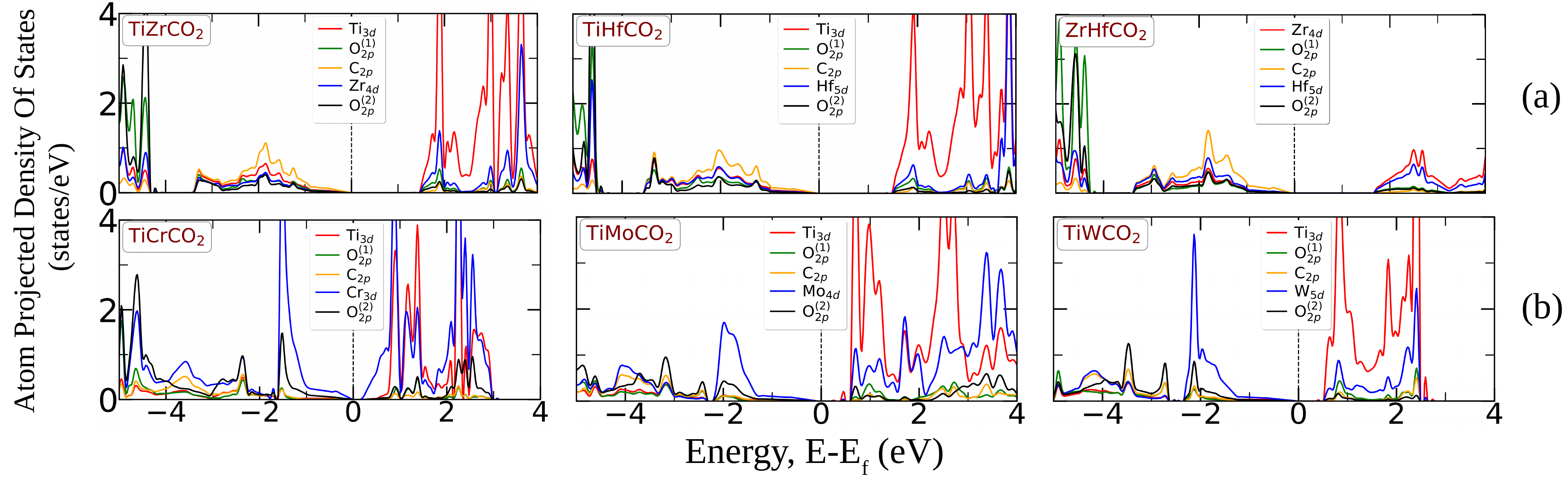}
		\caption{Atom projected densities of states of a) MM$^{\prime}$CO\textsubscript{2}, M,M$^{\prime}$= Ti, Zr, Hf; b) TiM$^{\prime}$CO\textsubscript{2}, M$^{\prime}$= Cr, Mo \& W respectively. Fermi energy in each case is set to 0 eV.}
		\label{fig5}
	\end{figure*}

	Janus MXenes can also be realised by replacing one of the M surfaces of M$_{2}$C MXene with a different transition metal M$^{\prime}$. Since all four M$_{2}$CO$_{2}$ MXenes where M is Sc, Ti, Zr, and Hf, are semiconductors, in this sub-section, we look into the structural properties and electronic ground states of MM$^{\prime}$CO$_{2}$ Janus where 6 different compounds are obtained from combinations of these four transition metal elements. 
	All six compounds in this series stabilise in Model 1 structure (Table \ref{table3}). For the three compounds made up of Ti, Zr and Hf as the transition metal components, both the O and the C get their necessary electrons from the transition metal constituents. Hence, Model 1 is the natural choice. For the three compounds where M is Sc, in spite of the shortage of 1 electron than those required by the anions, the ground state configuration is still Model 1. The reason can be explained the following way: ScM$^{\prime}$O$_{2}$ compounds can be thought to be derived from Sc$_{2}$CO$_{2}$ by replacing Sc$^{(2)}$ by Ti, Zr and Hf, all three having necessary electrons to fulfill the requirements of M$^{\prime}$-C and M$^{\prime}$-O$^{(2)}$ bonds. As a result, unlike in Sc$_{2}$CO$_{2}$, the O$^{(2)}$ prefers to occupy position A (Figure \ref{fig2} b)). However, the electronic ground states of Sc-based compounds are different from the other three. The Sc-based Janus are metals, while the others are semiconductors. The difference can be understood from the differences in their atom-projected densities of states (Figure \ref{fig5}a) and Figure S3, supplementary information). The common feature in the electronic structure of all compounds in this series is that states near the top of the valence band are mostly made up of the $d$ states of M$^{\prime}$ and $p$ states of C. In the Sc-based compounds, these states are delocalised near the Fermi level and extend into the unoccupied part of the spectrum. This is an artefact of a single unpaired electron present in C due to a lack of enough number of electrons in Sc. Since in the non-Sc MXenes, there are no such unpaired electrons, the hybridised states near the top of the valence bands are slightly deeper in energy, making these three compounds semiconducting.
	\begin{table*}[htb!]
		\small
		\caption{\bf{The structural parameters and electronic ground states of MM$^{\prime}$CO\textsubscript{2} (M=Sc, Ti, Zr, Hf; M$^{\prime}$=Ti, Zr, Hf, Cr, Mo, W) MXenes}}
		\label{table3}
		\begin{tabular*}{\textwidth}{@{\extracolsep{\fill}}ccccccc}
			\hline
			\multicolumn{1}{c}{\textbf{Systems}} & \multicolumn{1}{c}{\textbf{Electronic}} & \multicolumn{1}{c}{\textbf{Stable}} & \multicolumn{1}{c}{\textbf{Lattice-Constant}} & \multicolumn{2}{c}{\textbf{Bond-Lengths }} & \multicolumn{1}{c}{\textbf{Thickness}} \\
			\cline{5-6}
			\bf{(MM$^{\prime}$CO\textsubscript{2})} &\bf{ground state} & \bf{Configuration} & \bf{a=b}& \textbf{M-O} & \textbf{M$^{\prime}$-O} & \bf{t}\\
			& & & (\AA) & (\AA) & (\AA) & (\AA) \\
			\hline
			ScTiCO\textsubscript{2} & Metal & Model 1 & 3.15 & 1.99 & 2.01 & 4.55\\
			ScZrCO\textsubscript{2} & Metal & Model 1 & 3.30 &2.04 &2.12 & 4.61\\
			ScHfCO\textsubscript{2} & Metal & Model 1 & 3.27 & 2.03 & 2.10 & 4.62 \\
			TiZrCO\textsubscript{2} & Semiconductor & Model 1 & 3.19 & 2.02 & 2.09 & 4.52\\
			TiHfCO\textsubscript{2} & Semiconductor & Model 1 & 3.16 & 2.01 & 2.07 & 4.51\\
			ZrHfCO\textsubscript{2} & Semiconductor & Model 1 & 3.29 & 2.11 & 2.11 & 4.62\\
			\\
			TiCrCO\textsubscript{2} & Semiconductor & Model 3 & 2.87 & 1.94 & 1.99 & 4.61 \\
			TiMoCO\textsubscript{2} & Semiconductor & Model 3 & 2.95 & 1.96 & 2.08 & 4.77\\
			TiWCO\textsubscript{2} & Semiconductor & Model 3 & 2.96 & 1.96 & 3.63 & 4.79\\
			ZrCrCO\textsubscript{2} & Metal & Model 3 & 3.05 & 2.07 & 3.68 & 4.72 \\
			ZrMoCO\textsubscript{2} & Semiconductor & Model 3 & 3.10 & 2.07 & 3.77 & 4.90\\
			ZrWCO\textsubscript{2} & Semiconductor & Model 3 & 3.09 & 2.08 & 3.77 & 4.92\\
			HfCrCO\textsubscript{2} & Metal & Model 3 & 3.02 & 2.05 & 3.65 & 4.70 \\
			HfMoCO\textsubscript{2} & Semiconductor & Model 3 & 3.08 & 2.06 & 3.74 & 4.89\\
			HfWCO\textsubscript{2} & Semiconductor & Model 3 & 3.07 & 2.06 & 3.75 & 4.90\\
			\hline
		\end{tabular*}
	\end{table*}
	\subsubsection{The series MM$^{\prime}$CO\textsubscript{2}, M=Ti, Zr, Hf; M$^{\prime}$= Cr, Mo, W}
	The MM$^{\prime}$CO$_{2}$ series considered in this sub-section consists of a group IV element M and a group VI element M$^{\prime}$. Janus MXenes with such combinations are considered because M and M$^{\prime}$ elements considered here have significant differences in their electronegativity, a minimum of 7\%(Ti and Cr) to a maximum of 81\%(Hf and W). When M and M$^{\prime}$ are on two different surfaces, this difference in electronegativity can generate the internal electric field necessary for band bending. All nine Janus MXenes in this series stabilise in Model 3 configuration (Table \ref{table3}). In these compounds, there remain two excess electrons belonging to M$^{\prime}$ element after fulfilling requirements of C and O on either surface. The O atom associated with the M$^{\prime}$ surface still prefers to occupy C hollow sites as the large electron density of M$^{\prime}$ repeals O and makes site B unsuitable for it to occupy. Very large M$^{\prime}$-O bond distances as compared to M-O bond distances (Table \ref{table3}) support this.
	\begin{figure*}[htb!]
		\centering
		\includegraphics[scale=0.17]{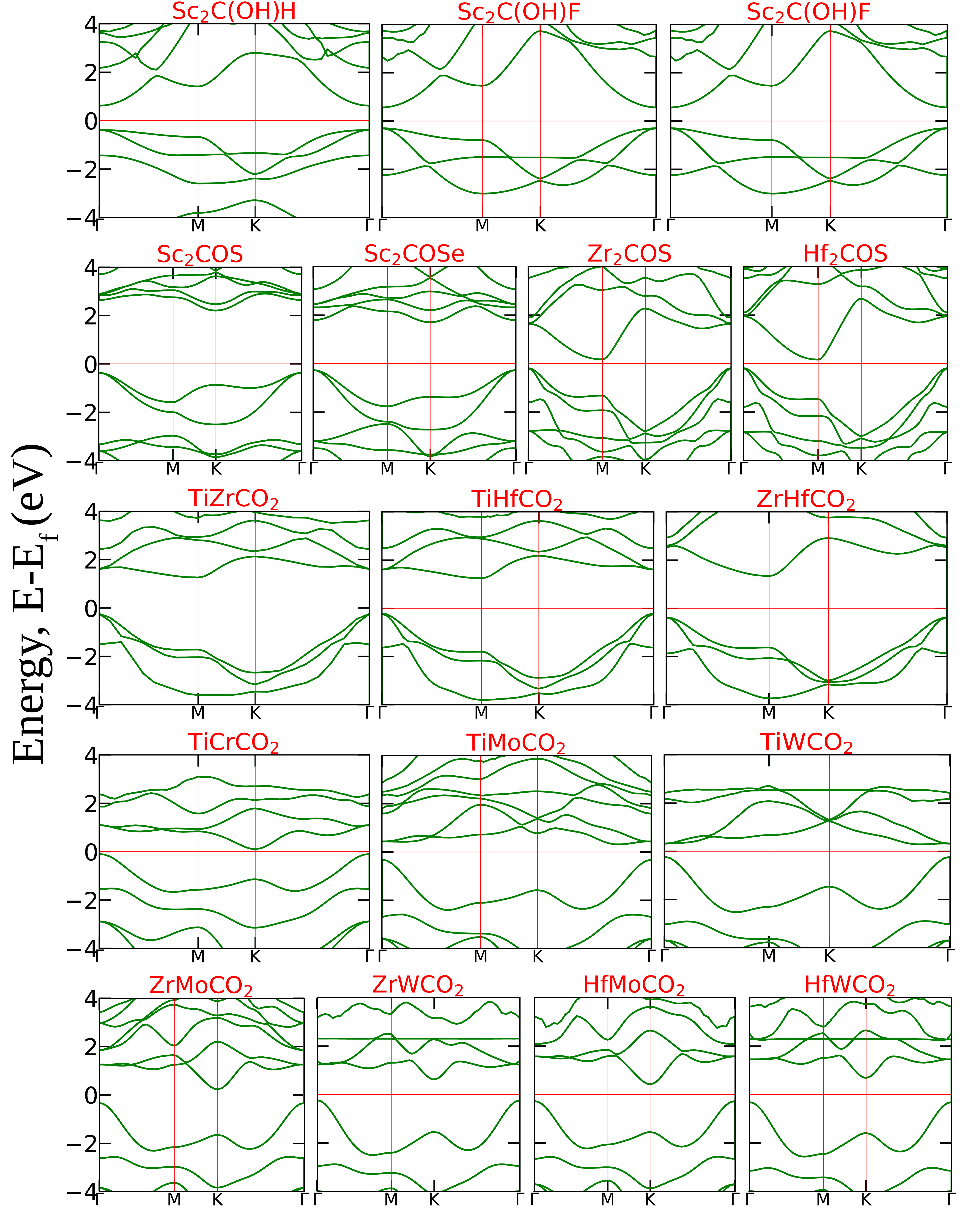}
		\caption{ Band Structures of 17 Janus MXenes that are potential photocatalysts for water splitting.}
		\label{fig6}
	\end{figure*}
	Atom projected densities of states of these compounds are shown in Figure \ref{fig5}b) and Figure S4, supplementary information. In all compounds, the states near the valence band edge are dominated by M$^{\prime}$-$d$ and O$^{(2)}$-$p$ states while states near conduction band edge are overwhelmingly contributed by hybridised M-M$^{\prime} d$ states. Out of the nine compounds, only two, ZrCrO$_{2}$ and HfCrO$_{2}$ are semi-metals, while TiCrO$_{2}$ is a semiconductor with a small band gap of only 0.2 eV. Upon inspecting these three compounds where Cr is the M$^{\prime}$ atom, we find that the electronic ground state is driven by Cr states in the unoccupied part. The semiconducting gap in TiCrO$_{2}$ has been possible due to localised Cr states in the unoccupied part; the localisation is due to strong hybridisation with Cr. In the two semi-metal compounds, the hybridisation is not that strong, and Cr states in the unoccupied part are more de-localised, creating states at the Fermi level. 
	
	\subsection{Band-Structures, Band edge alignments and Distribution of charge densities}
	Our calculations reveal 17 semiconductors among the 39 Janus MXene considered in this work. The next step for screening materials is to look at the band structures of these 17 to find out whether a compound is a direct or indirect band gap semiconductor and the size of its band gap. The first criterion is important to assess a compound's potential as a catalyst in the photo-splitting of water. The efficiency of the photocatalytic reaction will increase with less recombination of photo-generated electrons and holes. In the case of a semiconductor photocatalyst with an indirect band gap, the carriers cannot recombine as easily as in the case of a direct band gap semiconductor photocatalyst. The size of the band gap determines the part of the solar spectrum in which a given photocatalyst can operate. The band structures of these semiconducting Janus are shown in Figure \ref{fig6}. Except Sc$_{2}$C(OH)H and Sc$_{2}$C(OH)F, all other compounds have indirect band gaps. The size of their band gaps are given in Table \ref{table4}. Except 6 compounds, all others have a band gap of less than 1.23 eV. Therefore, these 11 compounds are potential photocatalysts that can absorb energy from the infrared part of the solar spectrum, while the rest can be active in the visible part of the spectrum. Our results agree very well with existing results on some of these compounds \cite{zhang_Sc, wang, huang,ozcan,hu,wong}.
	\begin{figure}[t]
		\includegraphics[scale=0.31]{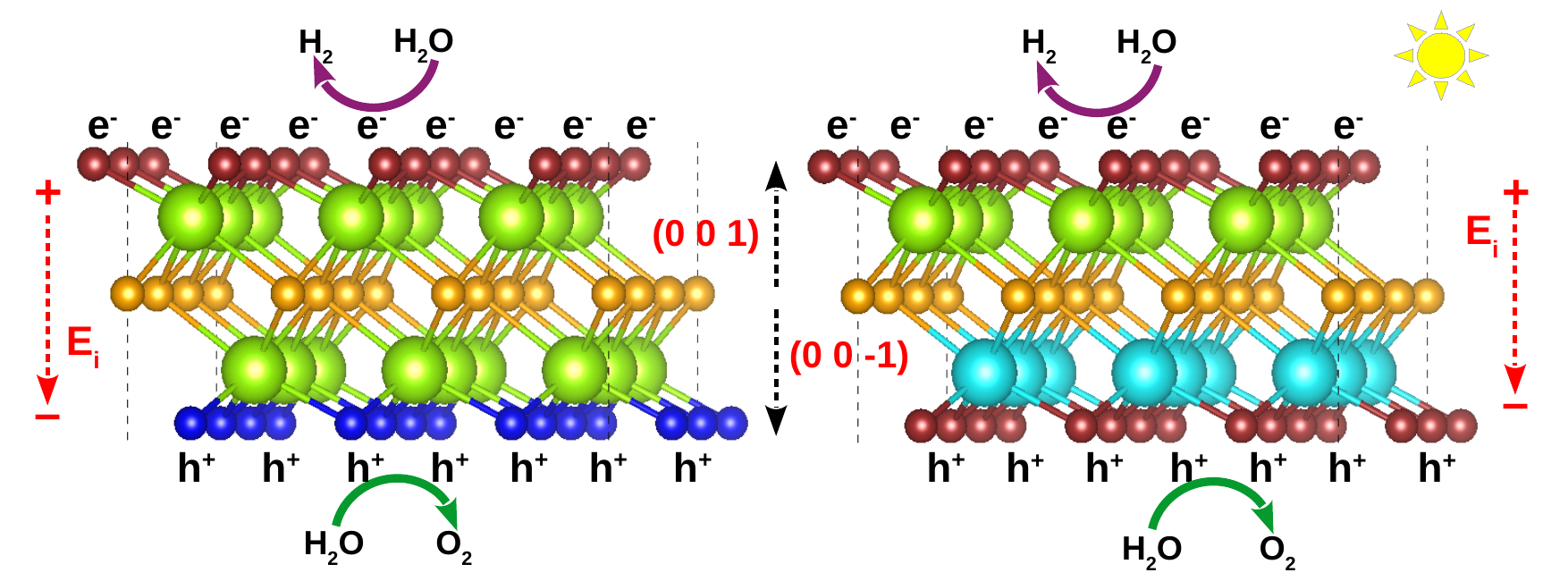}
		\caption{Schematic diagram of Photocatalytic water splitting  (HER \& OER) on the aymmetric surfaces of Janus MXenes. The figure on the left (right) is for M$_{2}$CTT$^{\prime}$(MM$^{\prime}$CT$_{2}$) MXene. }
		\hrule
		\label{fig7}
	\end{figure}
	\begin{figure*}[t]
		\includegraphics[scale=0.215]{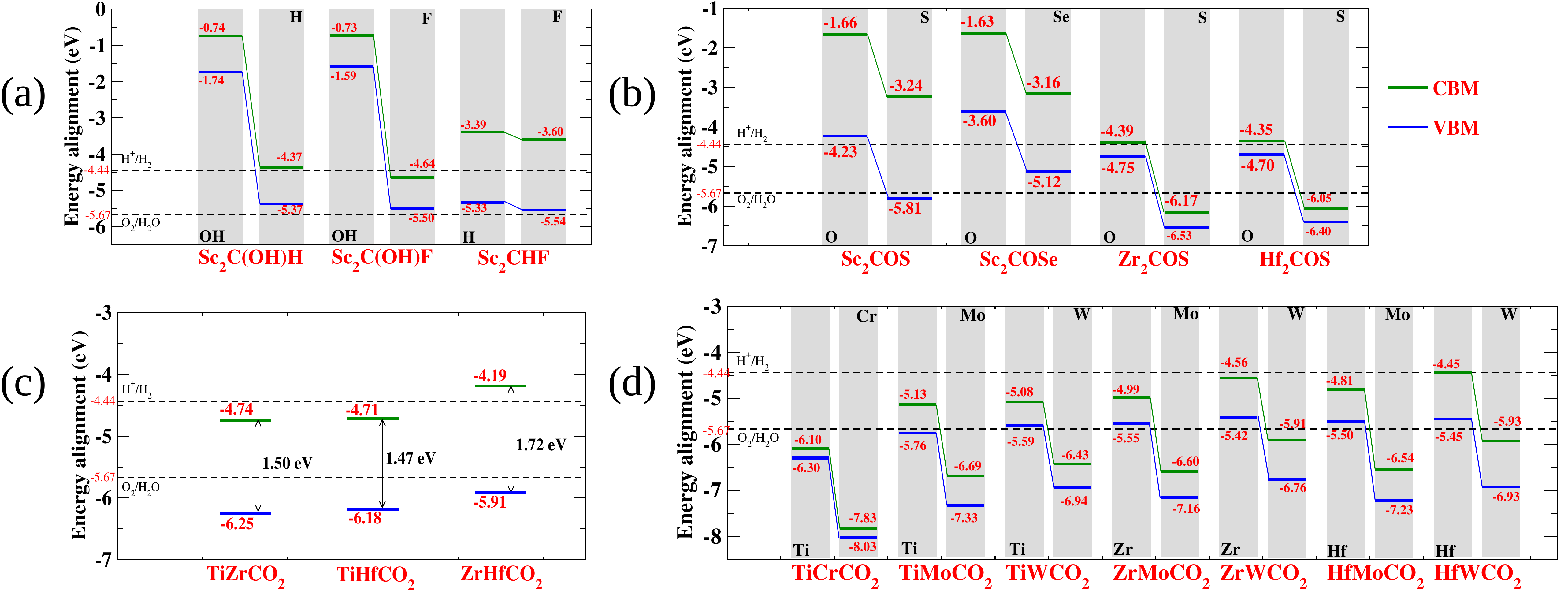}
		\caption{Re-adjusted band edges for the 17 Janus MXenes. The oxidation and reduction potentials for OER and HER are shown by black dotted lines. CBM(VBM) for each compound is shown by a green(blue) line. }
		\hrule
		\label{fig8}
	\end{figure*}
	The next step in our screening scheme is to find the electrostatic potential difference between the surfaces due to asymmetric charge distribution, an artefact of Janus structure, corresponding internal electric field, the intrinsic dipole moment generated across the surfaces, and re-alignment of band edges due to this potential difference. Once the semiconducting Janus MXenes with appropriate band alignments are found, surface contributions to their band decomposed charge densities are looked into. The internal electric field  $E_{i}$ and the potential difference between the surfaces $\Delta\Phi$, induced by the intrinsic dipole moment, $P$ are given as,
	\begin{eqnarray}
	E_{i} &=& \frac{\Delta\Phi}{et} \\
	\Delta\Phi &=& \frac{eP}{\epsilon A}
	\label{eq3}
	\end{eqnarray}
	where $\epsilon$ is the dielectric constant, $A$ is the area of the MXene surface, $t$  the thickness of the material and $e$ the electronic charge. $\Delta\Phi$ is calculated as the potential difference between the bottom surface (the surface at (00-1)) and the top surface (the surface at (001)). Calculated values of $t$ for the semiconducting Janus MXenes are given in Tables \ref{table1}-\ref{table3}. Results of $E_{i}, \Delta\Phi$ and $P$ are presented in Table \ref{table4}.
	\begin{table}[t]
		\small
		\caption{\bf{Band Gap (E$_g$), Electrostatic Potential Difference ($\Delta\Phi$) between two surfaces, Intrinsic Electric-Field (E$_i$), Dipole Moment (P), Over Potential ($\Delta$ E$_c$) between CBM and Water Reduction Potential \& Over Potential ($\Delta$ E$_v$) between VBM and Water Oxidation Potential for the Janus semiconductors considered in this work}}
		\label{table4}
		\begin{tabular*}{0.48\textwidth}{@{\extracolsep{\fill}}ccccccc}
			\hline
			\bf{System} & \bf{E$_g$} & \bf{$\Delta\Phi$} & \textbf{E$_i$}& \textbf{P} & \bf{$\Delta$ E$_c$} & \bf{$\Delta$ E$_v$} \\
			& (eV) & (eV) & (V/\AA) & (debye) & (eV) & (eV)\\
			\hline
			\\
			Sc\textsubscript{2}COHH & 1.00 & 3.64 & 0.63 & 0.91 & 3.70 & \bf{-0.30} \\
			Sc\textsubscript{2}COHF & 0.86 & 3.91 & 0.66 & 0.97 & 3.71 & \bf{-0.17} \\
			Sc\textsubscript{2}CHF & 1.94 & 0.21 & 0.04 & 0.05 & 1.05 & \bf{-0.13} \\
			\\
			Sc\textsubscript{2}COS & 2.57 & 1.58 & 0.39 & 0.48 & 2.78 & 0.14 \\
			Sc\textsubscript{2}COSe & 1.97 & 1.52 & 0.37 & 0.47 & 2.81 & \bf{-0.55} \\
			Zr\textsubscript{2}COS & 0.36 & 1.78 & 0.34 & 0.47 & 0.05 & 0.86 \\
			Hf\textsubscript{2}COS & 0.35 & 1.70 & 0.33 & 0.44 & 0.09 & 0.73 \\
			\\
			TiZrCO\textsubscript{2} & 1.50 & 0.0 & 0.0 & 0.0 & 0.58 & \bf{-0.30} \\
			TiHfCO\textsubscript{2} & 1.47 & 0.0 & 0.0 & 0.0 & 0.51 & \bf{-0.27} \\
			ZrHfCO\textsubscript{2} & 1.72 & 0.0 & 0.0 & 0.0 & 0.24 & 0.25 \\
			\\
			TiCrCO\textsubscript{2} & 0.20 & 1.73 & 0.38 & 0.33 &  \bf{-1.66} & 2.36 \\
			TiMoCO\textsubscript{2} & 0.64 & 1.57 & 0.33 &0.31  & \bf{-0.69} & 1.66 \\
			TiWCO\textsubscript{2} & 0.51 & 1.35 & 0.28 & 0.27 & \bf{-0.64} & 1.27 \\
			ZrMoCO\textsubscript{2} & 0.56 & 1.61 & 0.33 & 0.35 & \bf{-0.55} & 1.49 \\
			ZrWCO\textsubscript{2} & 0.86 & 1.35 & 0.27 & 0.30 & \bf{-0.12} & 1.09 \\
			HfMoCO\textsubscript{2} & 0.69 & 1.73 & 0.35 & 0.38 & \bf{-0.37} & 1.56\\
			HfWCO\textsubscript{2} & 1.00 & 1.47 & 0.30 & 0.32 & \bf{-0.01} & 1.26\\
			\hline
		\end{tabular*}
	\end{table}
	In cases of non-zero $P$ and $E_{i}$, we find $\Delta\Phi>0$. This means the top surface in these compounds is always at a lower potential. The profile of electrostatic potentials shown in Figure S5, supplementary information corroborates this. Internal electric field $E_{i}$, therefore, acts along negative $z$-direction. This will make the photo-generated electrons move from (00-1)surface to (001) surface. Consequently, the Oxygen evolution reaction (OER) and the hydrogen evolution reaction (HER) will prefer to occur on (0 0 -1) surface and (0 0 1) surface, respectively. This is schematically shown in Figure \ref{fig7}. The reduction and oxidation potentials are determined by the electrostatic potentials relative to the vacuum level. In the case of Janus MXenes, there are two different vacuum levels corresponding to different surfaces. Therefore, the shift in the reduction and oxidation potentials will be different for different surfaces. The band edges should be aligned accordingly. In Fig.\ref{fig8}, we show the re-adjusted positions of VBM and CBM on the two surfaces of the 17 Janus MXenes. The reduction(oxidation) overpotential $\Delta E_{c}$($\Delta E_{v}$) associated with the water splitting are given by
	\begin{eqnarray}
	\Delta E_{c}=\left\lbrace E_{CBM}-(-4.44)\right\rbrace  eV, \enskip \enskip  \Delta E_{v}=\left\lbrace -5.67-E_{VBM} \right\rbrace eV
	\end{eqnarray}
	$E_{CBM}$ and $E_{VBM}$ are the energies of the conduction band minima and the valence band maxima, respectively. Values of overpotentials for each compound are given in Table \ref{table4}. Positive values of $\Delta E_{c}$ and $\Delta E_{v}$ would imply alignment of CBM and VBM such that both reduction and oxidation processes can take place. From Table \ref{table4} we find that only four compounds, Sc$_{2}$COS, Zr$_{2}$COS, Hf$_{2}$COS and ZrHfCO$_{2}$ have both values positive. Six compounds, Sc$_{2}$C(OH)H, Sc$_{2}$C(OH)F, Sc$_{2}$CHF, Sc$_{2}$COS, TiZrCO$_{2}$,TiHfCO$_{2}$, have only positive $\Delta E_{c}$ implying that only the HER can take place. All seven MM$^{\prime}$CO$_{2}$ MXenes where M,M$^{\prime}$ are group-IV and group-VI elements, respectively, have only positive $\Delta E_{v}$. Consequently, only OER is possible in these compounds.
	\begin{figure*}[t]
		\includegraphics[scale=0.45]{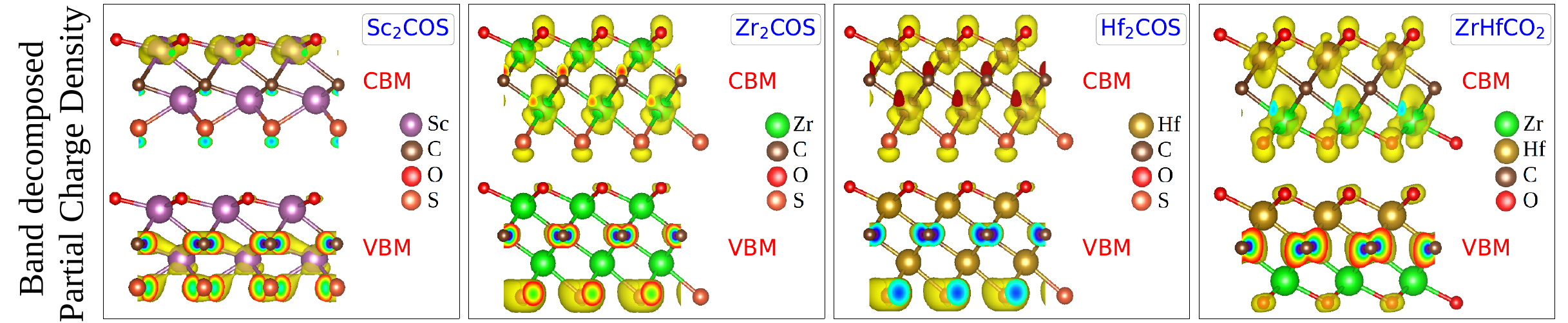}
		\caption{The Band decomposed Partial charge densities of constituents towards VBM and CBM of four Janus MXenes with proper band alignment are shown.}
		\hrule
		\label{fig9}
	\end{figure*}
	The effects of band bending due to the internal electric field and resulting re-positioning of VBM and CBM are prominent in the Janus MXenes considered (Figure \ref{fig8}). The larger the electric field, the larger is the bending. As a consequence, large changes are observed in cases of Sc$_{2}$C(OH)H and Sc$_{2}$C(OH)F. In both cases, the VBM is very close to the oxidation potential that would not have happened without the internal electric field. More importantly, in almost all cases, VBM and CBM, due to the band bending, lie on two different surfaces. Only exceptions are TiCrCO\textsubscript{2}, ZrMoCO\textsubscript{2}, ZrWCO\textsubscript{2}, HfMoCO\textsubscript{2} \& HfWCO\textsubscript{2} where both VBM and CBM are associated with the surface containing M$^{\prime}$ in these MM$^{\prime}$CO$_{2}$ MXenes (Figure S6, supplementary information). Though the overpotentials associated with HER in HfWCO$_{2}$ is only slightly negative, even a positive value would not have made it suitable for water splitting as the separation of charge carriers onto two surfaces will not be possible, enabling both HER and OER to take place simultaneously. No band bending is found in case of TiZrCO$_{2}$, TiHfCO$_{2}$ and ZrHfCO$_{2}$. For these compounds, $E_{i}$ is zero as there is no potential difference between the two surfaces (Figure S5, supplementary information). This is due to the fact that there is a little asymmetry between the two surfaces as the transition metal constituents are iso-electronic, have nearly the same electronegativity, and the chemical environment around them is nearly identical (Functional group occupies an identical position on both surfaces and bond lengths between the transition metal and oxygen associated with different surfaces are hardly different).
	
	The results obtained through the systematic screening finally identify four compounds, Sc$_{2}$COS, Zr$_{2}$COS, Hf$_{2}$COS, and ZrHfCO$_{2}$, as compounds suitable for photocatalytic water splitting. While Sc$_{2}$COS and ZrHfCO$_{2}$ can utilise the visible part of solar spectrum, Zr$_{2}$COS and Hf$_{2}$COS will be active in the Infra-red region. However, the last check for their competence as photocatalysts is done by examining their charge density isosurface distributions. Unless the VBM and the CBM states are contributed by components of two different surfaces, photocatalytic redox water-splitting cannot be feasible from the thermodynamic point of view. In Figure \ref{fig9}, we show the charge density distributions associated with VBM and CBM. The figures imply that contributions to VBM and CBM are coming from different surfaces. In Sc$_{2}$COS, VBM is made up of Sc$^{(2)}$ $d$, C $p$ and S $p$ states while CBM is mostly due to Sc$^{(1)}$ $d$ and O $p$ states. In Zr$_{2}$COS, VBM is due to S states while CBM is due to Zr$^{(1)}$-C hybridised states. A similar separation of contributions is observed in the other two compounds as well. 
	\section{Conclusions}
	A number of first-principles-based computational models to predict photocatalysts for water splitting in the MXene family far outnumber the experiments. However, almost all the works have investigated systems that are active in the visible region. Moreover, there is a lack of coherence between those studies as those were not done in a systematic way. In this work, we have tried to bridge that gap by adopting a combinatorial approach to generate 47 thinnest ($n=1$) MXenes with asymmetric surfaces that is Janus compounds and apply a screening criterion to systematically arrive at the ones that can host both OER and HER, preferably in the Infra-red part of the solar spectrum. Our calculations predict only 4 potential candidates, out of which 2 (Zr$_{2}$COS and Hf$_{2}$COS) can absorb the infra-red light while the other 2 (Sc$_{2}$COS and ZrHfCO$_{2}$) will be visible-active. As a by-product of our endeavor, we found 6(7) more Janus MXenes useful for water reduction(oxidation) reactions under Z-scheme photocatalysis. We have thoroughly investigated the role of structural properties and the bonding nature of the constituents to find clues that can help figure out predictors to screen materials. Our study suggests that the location of the functional group passivating MXene surfaces is connected with the electronic configuration of the constituents and resultant sharing of electrons between cations and anions. This, in turn, decides the ground state electronic property of a given Janus MXene. We also find that the re-alignment of bands due to the internal electric field would be possible if the two surfaces have elements or functional groups that have significantly different electronegativity. The electronic structures of all 47 compounds imply that a few more Janus MXenes with the semiconducting ground state can be realised by the application of strains as they are semi-metallic in nature. Almost all such compounds are M$_{2}$CTT$^{\prime}$, which implies that asymmetry in the functional group is more useful in exploiting the flexibility in structure-property relations in this family. This study provides important insights for experimentalists looking to tap the potential of MXenes in photocatalytic water splitting. The detailed and systematic approach adopted in our work can also be used for a more elaborate investigation of Janus MXenes.
	\section*{Acknowledgement}
	The authors gratefully acknowledge the Department of Science and Technology, India, for the computational facilities under Grant No. SR/FST/P-II/020/2009 and IIT Guwahati for the PARAM supercomputing facility where all computations are performed. The authors would like to thank Dr. Mukul Kabir for useful discussions.\\
	\section*{Author Information}
	Email address :\\
	\href{mailto:swatishaw@iitg.ac.in}{$\ast$swatishaw@iitg.ac.in}\\
	\href{mailto:subhra@iitg.ac.in}{$\dagger$subhra@iitg.ac.in}
	
\end{document}